\documentclass[reqno,a4paper,11pt]{article}
\pdfoutput=1
\usepackage[linktocpage=true]{hyperref}
\usepackage{color}

\usepackage{graphicx}
\usepackage[textwidth = 430 pt, textheight = 630 pt]{geometry}

\definecolor{MyDarkBlue}{rgb}{0.15,0.25,0.45}
\usepackage{epsfig,rotating}
\usepackage{amsmath,amssymb}
\usepackage{amsfonts}
\usepackage{mathrsfs}
\usepackage{bbm}

\usepackage{latexsym}
\usepackage{amsthm}

\usepackage[utf8x]{inputenc}

\usepackage{hyperref}
\hypersetup{
colorlinks=true,
citecolor=MyDarkBlue,
linkcolor=MyDarkBlue,
urlcolor=MyDarkBlue,
pdfauthor={Sam Palmer and Christian S\"amann},
pdftitle={The ABJM Model is a Higher Gauge Theory},
pdfsubject={hep-th}
breaklinks=true
}

\usepackage{tikz}
\usepackage{mathtools}

\theoremstyle{remark}

\theoremstyle{definition}

\theoremstyle{remark}

\let\fn\footnote
\renewcommand{\footnote}[1]{\linespread{1.1}\fn{#1}\linespread{1.29}}

\makeatletter\renewcommand{\section}{\@startsection
{section}{1}{\z@}{-3.5ex plus -1ex minus
    -.2ex}{2.3ex plus .2ex}{\bf }}
\makeatletter\renewcommand{\subsection}{\@startsection{subsection}{2}{\z@}{-3.25ex
plus -1ex minus
   -.2ex}{1.5ex plus .2ex}{\it }}
\makeatletter\renewcommand{\subsubsection}{\@startsection{subsubsection}{3}{-2.45ex}{-3.25ex
plus -1ex minus -.2ex}{1.5ex plus .2ex}{\it }}

\renewcommand{\@seccntformat}[1]{\@nameuse{the#1}.~~}

\makeatletter \@addtoreset{equation}{section}

\usepackage[toc,page]{appendix}

\def\slasha#1{\setbox0=\hbox{$#1$}#1\hskip-\wd0\hbox to\wd0{\hss\sl/\/\hss}}

\def\periodb#1{\setbox0=\hbox{$#1$}#1\hskip-\wd0\hbox to\wd0{-}}

\newcommand{\unit}{\mathbbm{1}}   			% identity map/matrix
\newcommand{\CF}{\mathcal{F}}

\newcommand{\CH}{\mathcal{H}}

\newcommand{\CN}{\mathcal{N}}

\newcommand{\fra}{\mathfrak{a}}				% frak-letters
\newcommand{\frg}{\mathfrak{g}}				% frak-letters
\newcommand{\frh}{\mathfrak{h}}				% frak-letters
\newcommand{\frm}{\mathfrak{m}}
\newcommand{\frder}{\mathfrak{der}}				% frak-letters

\newcommand{\frl}{\mathfrak{l}}

\newcommand{\FR}{\mathbbm{R}}     			% field of real numbers
\newcommand{\FC}{\mathbbm{C}}     			% field of complex numbers
\newcommand{\RZ}{\mathbbm{Z}}     			% ring of integers
\newcommand{\dd}{\mathrm{d}}     			% total differential
\newcommand{\di}{\mathrm{i}}     			% imaginary unit
\newcommand{\eps}{{\varepsilon}}			% antisymmetric tensors
\newcommand{\eand}{{\qquad\mbox{and}\qquad}}     		% and etc. in equations

\newcommand{\tr}{\,\mathrm{tr}\,}     			% trace
\newcommand{\agl}{\mathfrak{gl}}     			% algebras

\newcommand{\au}{\mathfrak{u}}
\newcommand{\asu}{\mathfrak{su}}

\newcommand{\sEnd}{\mathsf{End}\,}
\newcommand{\sHom}{\mathsf{Hom}\,}

\newcommand{\acton}{\vartriangleright}     			% span
\def\tyng(#1){\hbox{\tiny$\yng(#1)$}}			% small Young diagram
\def\tyoung(#1){\hbox{\tiny$\young(#1)$}}			% small Young diagram
\newcommand{\beq}{\begin{eqnarray}}
\newcommand{\eeq}{\end{eqnarray}}

\newcommand{\sft}{{\sf t}}

\begin{document}

\begin{titlepage}
\begin{flushright}
 EMPG--13--18
\end{flushright}
\vskip 2.0cm
\begin{center}
{\LARGE \bf The ABJM Model is a Higher Gauge Theory}
\vskip 1.5cm
{\Large Sam Palmer and Christian S\"amann}
\setcounter{footnote}{0}
\renewcommand{\thefootnote}{\arabic{thefootnote}}
\vskip 1cm
{\em Maxwell Institute for Mathematical Sciences\\
Department of Mathematics, Heriot-Watt University\\
Colin Maclaurin Building, Riccarton, Edinburgh EH14 4AS, U.K.}\\[0.5cm]
{Email: {\ttfamily sap2@hw.ac.uk~,~c.saemann@hw.ac.uk}}
\end{center}
\vskip 1.0cm
\begin{center}
{\bf Abstract}
\end{center}
\begin{quote}
M2-branes couple to a 3-form potential, which suggests that their description involves a non-abelian 2-gerbe or, equivalently, a principal 3-bundle. We show that current M2-brane models fit this expectation: they can be reformulated as higher gauge theories on such categorified bundles. We thus add to the still very sparse list of physically interesting higher gauge theories.
\end{quote}
\end{titlepage}

\section{Introduction and results}

M5-branes interact via M2-branes ending on them. An effective description of M5-branes should therefore be a gauge theory describing the parallel transport of the one-dimensional boundaries of these M2-branes in the worldvolume of the M5-branes. This is where higher gauge theory \cite{Baez:2004in,Baez:2010ya} enters the picture. In general, higher gauge theory with principal $n$-bundles captures the parallel transport of $(n-1)$-dimensional objects. 

It is known that the effective dynamics of a single M5-brane involves an $\CN=(2,0)$ tensor multiplet in six dimensions, which contains a 2-form potential $B$. Higher gauge theory naturally contains this 2-form potential, even in a non-abelian generalization: it is the gauge potential for the parallel transport of a one-dimensional object along a surface.

A Nahm transform is expected to connect the BPS sectors of effective descriptions of M2- and M5-branes and on loop space, such a transform was developed in \cite{Saemann:2010cp}. This suggests that M2-brane models should also have a higher gauge theoretic formulation. A first step in this direction was made in \cite{Palmer:2012ya}, where we showed that the 3-algebras underlying the Bagger-Lambert-Gustavsson (BLG) model \cite{Bagger:2007jr} and the Aharony-Bergman-Jafferis-Maldacena (ABJM) model \cite{Aharony:2008ug} are differential crossed modules. These differential crossed modules replace the notion of a gauge algebra in higher gauge theory with principal 2-bundles.

An important question remained open in \cite{Palmer:2012ya}. In a higher gauge theory, the so-called fake curvature should vanish and it was not clear how to achieve this. In this letter, we solve this issue and show that the ABJM model (and therefore also the BLG model) is a higher gauge theory based on principal 3-bundles rather than principal 2-bundles.

This further categorification is motivated as follows: A vanishing fake curvature $\CF$ for the specific differential crossed modules found in \cite{Palmer:2012ya} requires the usual 2-form curvature $F$ to vanish. This is clearly too strong a condition in the ABJM model. Here, we use the observation that a higher gauge theory on a principal 2-bundle with non-vanishing $\CF$ can be reformulated as a higher gauge theory on a principal 3-bundle for which the fake curvature does vanish \cite{Schreiber:N01,Baez:2010ya}. 

Additional motivation for the use of principal 3-bundles comes from recently constructed M5-brane models. They either make direct use of principal 3-bundles, as in the twistor construction of \cite{Saemann:2013pca}, or, as in the case of the (1,0) superconformal theories of \cite{Samtleben:2011fj}, can be reformulated in terms of principal 3- or 4-bundles \cite{Palmer:2013pka}. 

Finally, as mentioned in the abstract, M2-branes couple to a 3-form potential, which suggests an underlying picture involving principal 3-bundles.

Our reformulation of the ABJM model as a higher gauge theory also exhibits another interesting feature. In many cases, the gauge transformations in higher gauge theory are so general, that the theory can be gauge fixed to an abelian or even a trivial theory. This happens for example in \cite{Baez:2012bn}, where teleparallel gravity was reformulated as a higher gauge theory. All configurations there were a priori gauge equivalent to the trivial configuration. However, the Lagrangian of the underlying theory broke the usual higher gauge symmetry, allowing for non-trivial configurations. We show that the same happens in the case of the ABJM model. 

So far, very few examples of physically interesting higher gauge theories have been found. Most prominent amongst these are teleparallel gravity \cite{Baez:2012bn} and the BF-models as studied e.g.\ in \cite{Martins:2010ry}. Our reformulation of the ABJM model adds another example to this very short list.

\section{Higher gauge algebras}

\subsection{From hermitian 3-Lie algebras to differential crossed modules}

We start by briefly reviewing one of the results of \cite{Palmer:2012ya}, where we showed that hermitian 3-Lie algebras\footnote{as well as their real relatives, the (generalized) 3-Lie algebras.} are so-called differential crossed modules, which, in higher category theoretical terms, correspond to strict Lie 2-algebras. We stress here that $n$-Lie algebras are different from the categorifications of Lie algebras leading to Lie $n$-algebras. The latter appear as structure Lie $n$-algebras in principal $n$-bundles.

A {\em hermitian 3-Lie algebra} \cite{Bagger:2008se} is a complex vector space $\fra$ endowed with a bilinear-antilinear triple product $[-,-;-]:\fra\times \fra\times \fra\rightarrow \fra$ such that the hermitian fundamental identity
\begin{equation}\label{eq:fundamental_identity}
[[c,d;e],a;b]-[[c,a;b],d;e]=[c,[d,a;b];e]-[c,d;[e,b;a]]
\end{equation}
is satisfied for all $a,b,c,d,e\in\fra$. A hermitian 3-Lie algebra is {\em metric}, if it comes with a positive-definite Hermitian pairing $(-,-):\fra\times \fra\rightarrow \FC$ invariant in the sense that
\begin{equation}
 (d,[a,b;c])-([d,c;b],a)=0
\end{equation}
for all $a,b,c,d\in\fra$. 

We also define the maps $D:\fra\times \fra\rightarrow \sEnd(\fra)$ taking two elements of $\fra$ into an endomorphism of $\fra$ according to
\begin{equation}
 D(a;b)\acton c:=[c,a;b]~,
\end{equation}
where $a,b,c\in \fra$. Because of \eqref{eq:fundamental_identity}, the span of these $D(a;b)$ forms indeed a complex Lie algebra, which we call the {\em Lie algebra of inner derivations of $\fra$}, $\frder(\fra)$. One can impose reality conditions on the $D(a;b)$, e.g.\ by combining them into $\tilde{D}(a;b)=D(a;b)-D(b;a)$, cf.\ \cite{deMedeiros:2008zh}.

A particularly important example of a one-parameter family of metric hermitian 3-Lie algebra \cite{Bagger:2008se} is the case where $\fra\cong \agl(N,\FC)$ as a vector space,
\begin{equation}\label{eq:hermitian_3_Lie_bracket}
 [a,b;c]:=\kappa(ac^\dagger b-b c^\dagger a)\eand (a,b):=\tr(a^\dagger b)
\end{equation}
for $a,b,c\in\fra$ and $\kappa\in \FR$. The (real) Lie algebra of inner derivations, which is the span of the $\tilde{D}(a;b)$, is given by $\au(N)\times \au(N)$. This example underlies the ABJM model\footnote{This is actually the ABJM model with complexified matter fields, cf.\ \cite{Bagger:2008se}. For real matter fields, one reduces $\agl(N,\FC)$ to $\asu(N)\oplus \di\,\au(1)$, i.e.\ matrices in $\agl(N,\FC)$, whose traceless part is antihermitian and whose trace part is hermitian. Because we prefer not to clutter our discussion with the related technical details, we will work with complexified matter fields.} \cite{Aharony:2008ug}. 

As shown in \cite{deMedeiros:2008zh}, hermitian 3-Lie algebras can be derived from a metric Lie algebra $\frg=\frder(\fra)$ and a faithful complex unitary representation $\fra$. Such representations form special cases of differential crossed modules.

A {\em differential crossed module} is a pair of Lie algebras $\frg,\frh$ together with a map $\sft:\frh\rightarrow \frg$ and an action $\acton$ of $\frg$ onto $\frh$. We demand that $\sft$ is equivariant and that the so-called {\em Peiffer identity} holds:
\begin{equation}
 \sft(g\acton h)=[g,\sft(h)]\eand \sft(h_1)\acton h_2=[h_1,h_2]
\end{equation}
for all $g\in\frg$ and $h,h_1,h_2\in\frh$. A straightforward example of a differential crossed module is the differential crossed module of inner derivations of a Lie algebra $\frg$. Here, we put $\frh=\frder(\frg)\cong \frg$, $\acton$ is the adjoint action and $\sft$ is the identity.

Clearly, the above example of a hermitian 3-Lie algebra is such a differential crossed module: We put $\frh=\agl(N,\FC)$, regarded as an abelian Lie algebra, i.e.\ as a vector space with trivial Lie bracket. Moreover, $\frg=\au(N)\times \au(N)=\au(N)_L\times \au(N)_R$, $\sft$ is trivial and $\acton$ is given by the left- and right-product\footnote{i.e.\ the obvious matrix product} of elements of $\au(N)_L$ and $\au(N)_R$, respectively. We denote this differential crossed module by $\frm_{\rm ABJM}(N)$.

\subsection{Inner derivation 2-crossed modules}

Just as a Lie algebra comes with a differential crossed module governing the action of inner derivations, a differential crossed module (or strict Lie 2-algebra) comes with a differential 2-crossed module of inner derivations\footnote{Note that more generally, the Chevalley-Eilenberg algebra of the inner derivations of an $L_\infty$-algebra $\frg_\infty$ is known as the {\em Weil algebra} of $\frg_\infty$.} as implied e.g.\ by the results of \cite{Roberts:0708.1741}. In higher category theoretical terms, differential 2-crossed modules are certain Lie 3-algebras, which must not be confused with 3-Lie algebras. 

Recall that a differential 2-crossed module \cite{Conduche:1984:155} is a triple of Lie algebras $\frl,\frh,\frg$ arranged in a normal complex
\begin{equation}
 \frl\ \xrightarrow{~\sft~}\ \frh\ \xrightarrow{~\sft~}\ \frg~.
\end{equation}
There are $\frg$-actions $\acton$ onto $\frh$ and $\frl$ by derivations. The Peiffer identity $\sft(h_1)\acton h_2=[h_1,h_2]$ is now lifted by a $\frg$-equivariant bilinear map, called {\em Peiffer lifting} and denoted by $\{-,-\}: \frh\times \frh\rightarrow \frl$. These maps satisfy the following axioms for all $g\in\frg$, $h,h_1,h_2,h_3\in\frh$ and $\ell,\ell_1,\ell_2\in\frl$:
\begin{itemize}
 \setlength{\itemsep}{-1mm}
 \item[(i)] $\sft(g\acton \ell)=g\acton\sft(\ell)$ and $\sft(g\acton h)=[g,\sft(h)]$.
 \item[(ii)] $\sft(\{h_1,h_2\})=[h_1,h_2]-\sft(h_1)\acton h_2$.
 \item[(iii)] $\{\sft(\ell_1),\sft(\ell_2)\}=[\ell_1,\ell_2]$.
 \item[(iv)] $\{[h_1,h_2],h_3\}=\sft(h_1)\acton\{h_2,h_3\}+\{h_1,[h_2,h_3]\}-\sft(h_2)\acton\{h_1,h_3\}-\{h_2,[h_1,h_3]\}$.
 \item[(v)] $\{h_1,[h_2,h_3]\}=\{\sft(\{h_1,h_2\}),h_3\}-\{\sft(\{h_1,h_3\}),h_2\}$.
 \item[(vi)] $\{\sft(\ell),h\}+\{h,\sft(\ell)\}=-\sft(h)\acton \ell$.
\end{itemize}

Given a differential crossed module $\frh\xrightarrow{~\tilde{\sft}~}\frg$ with action $\tilde{\acton}:\frg\times \frh\rightarrow \frh$, the corresponding differential 2-crossed module of inner derivations, denoted $ \frder\big(\frh\xrightarrow{~\tilde{\sft}~}\frg\big)$, has the underlying normal complex \cite{Roberts:0708.1741}
\begin{equation}
\frh\ \xrightarrow{~\sft~}\ \frg \ltimes \frh\ \xrightarrow{~\sft~}\ \frg~.
\end{equation}
Recall that the Lie bracket on $\frg \ltimes \frh$ reads as 
\begin{equation}
[(g_1,h_1),(g_2,h_2)]:= ([g_1,g_2],[h_1,h_2]+g_1\tilde{\acton}h_2-g_2\tilde{\acton}h_1)~.
\end{equation}
The maps $\sft$ are defined as
\begin{equation}
 \sft(h):=(\tilde{\sft}(h),-h)\eand \sft(g,h):=\tilde{\sft}(h)+g~,
\end{equation}
the $\frg$-actions and the Lie bracket on $\frh$ are given by
\begin{equation}
g\acton h := g~\tilde{\acton}~h\eand g_1 \acton (g_2,h):=([g_1,g_2],g_1~\tilde{\acton}~h)
\end{equation}
and the Peiffer lifting reads as
\begin{equation}
 \{(g_1,h_1),(g_2,h_2)\}:=g_2\tilde{\acton} h_1
\end{equation}
for all $g,g_1,g_2\in\frg$, $h,h_1,h_2\in \frh$. One readily checks that this structure satisfies the axioms of a differential 2-crossed module.

\subsection{Inner derivations of $\frm_{\rm ABJM}(N)$}

The inner derivations of $\frm_{\rm ABJM}(N)$ are captured by a differential 2-crossed module that is constructed from $\frm_{\rm ABJM}(N)$ as described in the previous section. To simplify the discussion, let us use the following picture: We consider a chain complex of block matrices
\begin{equation}\label{eq:rep_d2cm}
\frh:=\left(\begin{array}{cc} 0 & \agl(N,\FC)\\ 0 & 0\end{array}\right)\ \xrightarrow{~\sft~}\ \frg\ltimes \frh:=\left(\begin{array}{cc} \au(N) & \agl(N,\FC)\\ 0 & \au(N)\end{array}\right)\ \xrightarrow{~\sft~}\ \frg:=\left(\begin{array}{cc} \au(N) & 0\\ 0 & \au(N)\end{array}\right)~,
\end{equation}
where the two maps $\sft:\frh \rightarrow \frg\ltimes \frh$ and $\sft:\frg\ltimes \frh\rightarrow\frg$ read as
\begin{equation}
\sft:\left(\begin{array}{cc} 0 & h\\ 0 & 0\end{array}\right)\mapsto \left(\begin{array}{cc} 0 & -h\\ 0 & 0\end{array}\right) \eand 
\sft:\left(\begin{array}{cc} g_L & h\\ 0 & g_R\end{array}\right)\mapsto \left(\begin{array}{cc} g_L & 0\\ 0 & g_R\end{array}\right)
\end{equation}
respectively, for $g_{L,R}\in\au(N)$ and $h\in\agl(N,\FC)$. All $\frg$-actions as well as the Lie algebra commutators are given by the corresponding matrix commutators. The Peiffer lifting is defined as
\begin{equation}
\left\{\left(\begin{array}{cc} g_{L1} & h_1\\ 0 & g_{R1}\end{array}\right),\left(\begin{array}{cc} g_{L2} & h_2\\ 0 & g_{R2}\end{array}\right)\right\}:=\left(\begin{array}{cc} 0 & g_{L2} h_1-h_1 g_{R2}\\ 0 & 0\end{array}\right)~,
\end{equation}
where $g_{L1,2},g_{R1,2}\in \au(N)$ and $h_{1,2}\in\agl(N,\FC)$. As a consistency check, one can easily verify that this Peiffer lifting indeed captures the failure of the Peiffer identity according to
\begin{equation}
 \sft(\{(g_1,h_1),(g_2,h_2)\})=[(g_1,h_1),(g_2,h_2)]-\sft(g_1,h_1)\acton (g_2,h_2)~.
\end{equation}
We will denote this differential 2-crossed module by $\frder(\frm_{\rm ABJM}(N))$.

\subsection{General higher gauge theory}

We will need the basics of the local description of higher gauge theory by a connective structure on a trivial principal 3-bundle over $M=\FR^{1,2}$. The detailed picture for gauge theory on principal 3-bundles was developed in \cite{Saemann:2013pca}, see \cite{Martins:2009aa} for a partial earlier account. Let us work for the moment with a general differential 2-crossed module $\frl\stackrel{\sft}{\rightarrow}\frh\stackrel{\sft}{\rightarrow}\frg$, we will restrict ourselves to the case $\frder(\frm_{\rm ABJM}(N))$ in the next section.

Consider 1-, 2- and 3-form potentials $A\in \Omega^1(M,\frg)$, $B\in \Omega^2(M,\frh)$ and $C\in \Omega^3(M,\frl)$. From these, we construct the corresponding field strengths
\begin{equation}
  F:=\dd A+\tfrac{1}{2}[A,A]~,~~~H:=\dd B+A\acton B~,~~~G:=\dd C+A\acton C+ \{B,B\}~.
\end{equation}

The gauge transformations of the gauge potentials are given by \cite{Saemann:2013pca}
\begin{equation}\label{eq:gauge_transformations}
 \begin{aligned}
\tilde C&=\gamma^{-1}\acton C-\tilde\nabla^0\big(\Sigma-\tfrac12\{\Lambda,\Lambda\}\big)+\{\tilde B,\Lambda\}+\{\Lambda,\tilde B\}-\{\Lambda,\tilde{\nabla}\Lambda+\tfrac12[\Lambda,\Lambda]\}~,\\
 \tilde B&=\gamma^{-1}\acton B-\tilde{\nabla}^0\Lambda-\tfrac12\sft(\Lambda)\acton \Lambda-\sft(\Sigma)~,\\
 \tilde A&=\gamma^{-1}A \gamma+\gamma^{-1}\dd \gamma-\sft(\Lambda)~,
 \end{aligned}
\end{equation}
where $\gamma$ is a function on $M$ taking values in a Lie group $\mathsf{G}$ with $\frg=\mathsf{Lie}(\mathsf{G})$, $\Lambda\in \Omega^1(M,\frh)$ and $\Sigma\in \Omega^2(M,\frl)$. Moreover, we used abbreviations $\tilde \nabla\ :=\ \dd+\tilde A\acton$ and $\tilde \nabla^0\ :=\ \dd+\big(\tilde A+\sft(\Lambda)\big)\acton$.

For the higher gauge theory to describe a parallel transport of membranes along three-dimensional volumes that is invariant under reparameterizations of the volume, the so-called {\em fake curvatures} have to vanish:
\begin{equation}\label{eq:fake_curvature_conditions}
 \CF:=F-\sft(B)=0\eand \CH:=H-\sft(C)=0~.
\end{equation}
Together with the Bianchi identity for $F$, $F-\sft(B)=0$ implies that $\sft(H)=0$.

Coupling matter fields to this gauge structure is not straightforward. Assume a matter field $\phi$ behaves as  $\tilde{\phi}:=g\acton \phi$ under gauge transformations. The corresponding covariant derivative then does {\em not} transform as $\widetilde{\nabla\phi}=g\acton (\nabla\phi)$ due to the modified transformation of the gauge potential $A$, cf.\ \eqref{eq:gauge_transformations}. There are essentially two solutions to this problem: First, we can impose additional conditions on the matter fields, such as e.g.\ $\sft(\phi)=0$, which is appropriate in the case of principal 2-bundles \cite{Saemann:2012uq,Saemann:2013pca}. The other possibility is to restrict to so-called {\em ample gauge transformations} with $\sft(\Lambda)=0$, cf.\ \cite{Palmer:2012ya}. This will turn out to be the appropriate condition for the ABJM model.

\section{The ABJM model}

\subsection{Higher gauge theoretic formulation of the ABJM model}

The ABJM model describes a stack of $N$ flat M2-branes with a $\FC^4/\RZ_k$ orbifold in the transverse directions. These eight transverse directions of the M2-branes are thus packaged into four complex fields $Z^A$, $A=1,\dots,4$, which have spinors $\psi^A$ as their superpartners. These matter fields take values in $\frh:=\agl(N,\FC)$. The gauge potential one-form $A$ lives in $\frg:=\au(N)\times\au(N)$. We use the representation \eqref{eq:rep_d2cm} of the differential 2-crossed module $\frder(\frm_{\rm ABJM}(N))$, where the action of the gauge potentials on matter fields corresponds to the matrix commutator. Besides this, there is also the ordinary matrix product between matter fields and their adjoints, which we will need for the potential terms in the ABJM model.

The ABJM action can then be written in the following way:
\begin{equation}
S_{\rm ABJM} = \int_{\FR^{1,2}} \tr\left(\tfrac{k}{4\pi
}\eta~A\wedge(\dd A
+\tfrac{1}{3}[A,A])-\nabla Z_A^{\dag}\wedge \star \nabla Z^A-\star\di\bar\psi^{A}\wedge \slasha{\nabla} \psi_A\right)+V~,
\end{equation}
where $\nabla=\dd+A\acton$ and $\eta=-\sigma_3\otimes \unit_N$ yields a metric of split signature on the gauge algebra $\au(N)\times \au(N)$. By $\tr(-)$, we mean the trace in the matrix representation \eqref{eq:rep_d2cm}. The potential is given by 
\begin{equation}
\begin{aligned}
V=\tfrac{2\pi}{k}& \int_{\FR^{1,2}}~\star{\rm tr}\Big(-\di\bar\psi^{A\dag} \psi_{A} Z^\dag_B
Z^B-\di\bar\psi^{A\dag} Z^B Z^\dag_B\psi_{A}+2\di\bar\psi^{A\dag}\psi_{B} Z_A^\dag Z^B-2\di\bar\psi^{A\dag} Z^B Z_A^\dag\psi_{B}\\
&\hspace{2.5cm}+\di\varepsilon_{ABCD}\bar\psi^{A\dag}
Z^C\psi^{B\dag} Z^D   -\di\varepsilon^{ABCD}Z_D^\dag\bar \psi_A Z_C^\dag\psi_B-\tfrac{4\pi^2}{3k}\Upsilon^{CD}_B\Upsilon^{\dagger B}_{CD}\Big)~,\\
\Upsilon^{CD}_B &:=
  Z^CZ^\dagger_B Z^D-\frac{1}{2}\delta^C_BZ^EZ^\dagger_E Z^D+\frac{1}{2}\delta^D_BZ^EZ^\dagger_E Z^C~.
\end{aligned}
\end{equation}
This theory exhibits $\CN=6$ supersymmetry and it has passed some highly non-trivial tests as an effective description of M2-branes. 

Next, we extend this action to implement the fake curvature conditions \eqref{eq:fake_curvature_conditions}, introducing 2- and 3-form potential $B\in \Omega^2(\FR^{1,2},\frg\ltimes \frh)$ and $C\in \Omega^2(\FR^{1,2},\frh)$. In the matrix representation \eqref{eq:rep_d2cm} of $\frder(\frm_{\rm ABJM}(N))$, the fake curvature conditions amount to 
\begin{equation}\label{eq:fake_curvature_conds_matrix}
\begin{aligned}
B=\left(\begin{array}{cc} F_L & b\\ 0 & F_R\end{array}\right)~,~~H= \left(\begin{array}{cc} 0 & \dd b +A_L b-bA_R\\ 0 & 0\end{array}\right)=\sft(C)=\left(\begin{array}{cc} 0 & -c\\ 0 & 0\end{array}\right)
\end{aligned}
\end{equation}
for some $b,c\in\agl(N,\FC)$, where $A_L$ and $A_R$ are the first and second block diagonal entries of $A$ and $F_{L,R}=\dd A_{L,R}+\tfrac{1}{2}[A_{L,R},A_{L,R}]$. Note that because of $\sft(H)=0$, $H$ has no block diagonal entries. 

To enforce \eqref{eq:fake_curvature_conds_matrix}, we introduce Lagrange multipliers $\lambda_{1}\in \Omega^1(\FR^{1,2},\frg)$, $\lambda_{2}\in \Omega^0(\FR^{1,2},\frg\ltimes \frh)$ and $\lambda_{3}\in \Omega^3(\FR^{1,2},\frg)$, adding the following terms to the action\footnote{As it stands, this action is not real. However, one can either impose reality conditions on $H$ and $\lambda_2$ or add complex conjugate terms to correct for this in a straightforward manner. Again we suppress these technical details.}:
\begin{equation}
 S_{\rm HGT}=S_{\rm ABJM}+ \int_{\FR^{1,2}}\tr\left(\lambda^\dagger_1\wedge(F-\sft(B))+\lambda_2^\dagger(H-\sft(C))+\lambda_3^\dagger\sft(\lambda_2)\right)~.
\end{equation}

Varying with respect to $\lambda_1$ and $\lambda_2$, we obtain
\begin{equation}\label{eq:variation_l1_l2}
 F-\sft(B)=0~,~~~H-\sft(C)+\sft^*(\lambda_3)=0~,
\end{equation}
where $\sft^*$ is the adjoint to $\sft$. This map is the trivial embedding of $\frg$ into $\frg\ltimes \frh$. Because $H-\sft(C)$ is a block off-diagonal in $\frg\ltimes \frh$, \eqref{eq:variation_l1_l2} reduces to
\begin{equation}
 F-\sft(B)=0~,~~~H-\sft(C)=0~,~~~\lambda_3=0~.
\end{equation}
Varying $S_{\rm HGT}$ with respect to $\lambda_3$ and $C$, we have
\begin{equation}
 \sft(\lambda_2)=\sft^*(\lambda_2)=0~~~\Leftrightarrow~~~\lambda_2=0~,
\end{equation}
where $\sft^*$ is here the obvious projection of $\frg\ltimes \frh$ onto $\frh$. Finally, varying the action with respect to $B$ yields
\begin{equation}
 \sft^*(\lambda_1)+\nabla\lambda_2=0~,
\end{equation}
which implies $\lambda_1=0$ due to $\lambda_2=0$.

Varying $S_{\rm HGT}$ with respect to the gauge potential, we obtain the usual equation of motion of the ABJM model plus terms containing the Lagrange multipliers $\lambda_1$ and $\lambda_2$. Since both vanish on-shell, we recover
\begin{equation}\label{eq:eom_F}
F=\star\left(\nabla Z^A Z^\dagger_A-Z^A\nabla  Z^{\dagger}_{A}+Z^\dagger_A\nabla Z^A -\nabla Z^\dagger_A Z^A  -\di \bar\psi^A \gamma\psi^\dagger_A-\di \bar\psi^{\dagger A} \gamma\psi_A\right)~,
\end{equation}
where $\gamma=\gamma_i\dd x^i$. The equations of motion for the matter fields remain obviously those of the ABJM model. Note that the four-form curvature $G$ trivially vanishes, as our trivial principal 3-bundle lives over $\FR^{1,2}$.

Altogether, the action $S_{\rm HGT}$ yields the equations of motion of the ABJM model, together with the fake curvature conditions \eqref{eq:fake_curvature_conds_matrix}. We therefore reformulated the ABJM model as a higher gauge theory. 

Supersymmetry and gauge symmetry of the ABJM model are trivially preserved, if we demand that $\lambda_{1,2,3}$ transform appropriately. Explicitly, we can demand that the fields $B$ and $C$ transform in the same way as $\sft^*(F)$ and $\sft^*(H)$, which renders the fake curvature conditions invariant under supersymmetry. The Lagrange multipliers can then be chosen to be invariant under supersymmetry, too. 

Gauge transformations should act on the Lagrange multipliers as
\begin{equation}
      \lambda_1\rightarrow \tilde{\lambda}_1=\gamma\lambda_1\gamma^{-1}+\gamma[\lambda_2,\Lambda^\dagger]\gamma^{-1}~,~~~\lambda_{2,3}\rightarrow \tilde{\lambda}_{2,3}=\gamma\lambda_{2,3}\gamma^{-1}~,
\end{equation}
where $\gamma\in \Omega^0(M,\mathsf{G})$ and $\Lambda\in \Omega^1(M,\frg\ltimes\frh)$ are the gauge parameters. The second term in the $\lambda_1$ transformation renders the action gauge invariant off-shell. The 2- and 3-form potentials $B$ and $C$ transform as specified in \eqref{eq:gauge_transformations}. 

Note however, that the ABJM model is {\em not} invariant under the general tensor transformations parametrized by $\Lambda$ in \eqref{eq:gauge_transformations}. In particular, the equation of motion for the 2-form curvature \eqref{eq:eom_F} breaks this symmetry. We are therefore left with the ample gauge transformations, which are parametrized by a $\Lambda$ with $\sft(\Lambda)=0$. This solves a common problem when working with higher gauge theories: In many cases, e.g.\ if $\sft:\frh\rightarrow \frg$ is surjective, the potential 1-form $A$ can be gauged away by a tensor transformation, leaving an abelian theory. This is not possible if these transformations are broken down to the ample ones. 

The same observation was made in \cite{Baez:2012bn}, where teleparallel gravity was reformulated as a higher gauge theory. Here, all field configurations can be gauge transformed away by tensor transformations. However, the action of the theory is not invariant under these symmetries, leaving only the usual group-valued gauge transformations.

The $\Sigma$-transformations in \eqref{eq:gauge_transformations} affect only the new terms added to $S_{\rm ABJM}$, which contain the Lagrange multipliers. All these terms are invariant under these transformations.

\subsection{ABJ-model}

The ABJ model \cite{Aharony:2008gk} is a Chern-Simons matter theory closely related to the ABJM model and also invariant under $\CN=6$ supersymmetry. We follow precisely the same formulation as above, merely replacing $\frm_{\rm ABJM}(N)$ by $\frm_{\rm ABJ}(N_1,N_2)$, which is the differential crossed module $\sHom(\FC^{N_2},\FC^{N_1})\stackrel{\sft}{\rightarrow}\au(N_1)\times \au(N_2)$. We then obtain a differential 2-crossed module of inner derivations, which we can represent in terms of matrices as
 \begin{equation}
\begin{aligned}
\left(\begin{array}{cc} 0 & \sHom(\FC^{N_2},\FC^{N_1})\\ 0 & 0\end{array}\right)\rightarrow \left(\begin{array}{cc} \au(N_1) & \sHom(\FC^{N_2},\FC^{N_1})\\ 0 & \au(N_2)\end{array}\right)\rightarrow \left(\begin{array}{cc} \au(N_1) & 0\\ 0 & \au(N_2)\end{array}\right)~.
\end{aligned}
\end{equation}

It does not seem possible to use more general types of differential crossed modules to obtain $\CN=6$ Chern-Simons matter theories. The hermitian 3-Lie algebras underlying such models seem to be very rigid. Note in particular that, as shown in \cite{Cherkis:2008ha}, the only hermitian 3-Lie brackets that can be written as products of matrices and their adjoints are of the form \eqref{eq:hermitian_3_Lie_bracket}.

\subsection{BLG model and generalizations}

We can restrict the ABJM model to the Bagger-Lambert-Gustavsson model \cite{Bagger:2007jr} by restricting to $\frm_{\rm ABJM}(2)$ and imposing a reality condition, reducing $(\agl(2,\FC),\au(2)\times \au(2))$ to $(\asu(2)\oplus \di\,\au(1),\asu(2)\times \asu(2))$. This turns the hermitian 3-Lie algebra into the (real) 3-Lie algebra $A_4$, which is a real four dimensional vector space with totally antisymmetric 3-bracket
\begin{equation}\label{eq:A4}
[e_\mu,e_\nu,e_\rho]=\eps_{\mu\nu\rho\sigma}e_\sigma~,
\end{equation}
on the basis elements $e_\mu\in A_4$. The Lie algebra of inner derivations is represented by the matrices
\begin{equation}
\begin{aligned}
\left(\begin{array}{cc} 0 & \asu(2)\oplus \di\,\au(1)\\ 0 & 0\end{array}\right)\rightarrow \left(\begin{array}{cc} \asu(2) & \asu(2)\oplus \di\,\au(1)\\ 0 & \asu(2)\end{array}\right)\rightarrow \left(\begin{array}{cc} \asu(2) & 0\\ 0 & \asu(2)\end{array}\right)~.
\end{aligned}
\end{equation}
The resulting action $S_{\rm HGT}$ will have enhanced $\CN=8$ supersymmetry. We can now reformulate this action in terms of 3-dimensional $\CN=2$ superfields, cf.\ \cite{Cherkis:2008ha}, and replace the above differential 2-crossed module with a more general one. For example, we can use an inner derivation differential 2-crossed modules of a differential crossed module arising from a generalized 3-Lie algebra as constructed in \cite{Palmer:2012ya}. The result is an $\CN=2$ supersymmetric Chern-Simons matter theory. If one is interested in such theories that are conformal, then one can take these generalizations and add further interaction terms to the potentials as discussed in \cite{Akerblom:2009gx}.

\section*{Acknowledgements}
We would like to thank Martin Wolf for discussions and very helpful comments on a first draft of this letter. This work was supported by the EPSRC Career Acceleration Fellowship EP/H00243X/1.

\end{document}